# Bulk Nanocrystalline Thermoelectrics Based on Bi-Sb-Te Solid Solution

[1]Bulat L.P., [2]Pshenai-Severin D.A., [3]Karatayev V.V.,
[3]Osvenskii V.B., [3,5]Parkhomenko Yu.N., [3]Lavrentev V., [3]Sorokin A.,
[4]Blank V.D., [4]Pivovarov G.I., [5]Bublik V.T., [5]Tabachkova N.Yu.

[1] *National Research University ITMO, St. Petersburg, Russia*
[2]*Ioffe Physical Technical Institute, St Petersburg, Russia*
[3]*GIREDMET Ltd., Moscow, Russia*
[4]*Technological Institute of Superhard and New Carbon Materials, Troitsk, Russia*
[5]*National University of Science and Technology "MISIS", Moscow, Russia*

### *Contents*

## 1. Introduction

Thermoelectric energy conversion represents one of ways of direct conversion of the thermal energy to the electric energy. The thermoelectric converters – a thermoelectric power generator or a thermoelectric cooler are solid-state devices, therefore it possesses following advantages: environmentally cleanliness, simplicity of management and convenience of designing, high reliability and possibility of longtime operation without service, absence of moving parts, absence of noise, vibration and electromagnetic noise, compactness of modules, independence of space orientation, ability to carry out of considerable mechanical overloads. Despite obvious advantages of thermoelectric conversion it has the important lack – rather small value of the efficiency η; in the best cases η = (5 – 8) %. Therefore thermoelectric generators are used today, as a rule, only in «small power», where it is impossible or is economically inexpedient to bring usual electric mains: for power supply of space missions, at gas and oil pipelines, for power supply of sea navigating systems, etc. Nevertheless, the thermoelectric method of utilization of waste heat from units of cars and vessels is unique technically possible. It appears the thermoelectric generators can save up to 7 % of automobile fuel. The thermoelectric coolers become economically justified at enough small cooling power Qc as a rule no more than 10 – 100 W. However the thermoelectric coolers are widely applied in the most different areas: domestic refrigerators, water-chillers, picnic-boxes; coolers for medicine and biology, for scientific and laboratory equipment, refrigeration systems for transport facilities. Very important area connects with strong up-to-date requirements for the thermal management of micro- and optoelectronics elements, including microprocessors and integrated circuit; the requirements have essentially increased owing to the increase in their speed and miniaturization. And the desired value of local heat removal from concrete spots of chips can be realized only by means of the thermoelectric cooling.

It is known, that the efficiency of thermoelectric generators and the coefficient of performance (COP) of thermoelectric coolers are defined by the dimensionless parameter of a thermoelectric material $ZT = \dfrac{\sigma\alpha^2}{\kappa}T$, where $T$ – is the absolute temperature, $Z$ – thermoelectric figure of merit, $\sigma$, $\kappa$, $\alpha$ – accordingly electric conductivity, thermal conductivity and Seebeck coefficient (thermoelectric power) of a used material.

Today the best commercial thermoelectric materials (thermoelectrics) has the efficiency $ZT$ =1.0. Let us underline that $ZT$ increased from 0.75 only to 1.0 during lust five decades. Obviously, competitiveness of thermoelectric generators and coolers will rise if it will be possible to increase the figure of merit. Thus the thermoelectric generation and cooling can provide a weighty contribution to a decision of the problem of utilization of renewed energy sources, a recycling of a low potential heat, and maintenance of storage of a foodstuff. Recently the increasing attention is involved to the thermoelectric refrigeration and the electric power generation as to environmentally clean methods. It is caused by several reasons. The main of them is caused by new scientific results on improve of the thermoelectric figure of merit. Important results in the development of highly effective nanostructured thermoelectric materials have been published last decade; see for example the reviews (Dresselhaus et al., 2007; Minnich et al., 2009; Dmitriev & Zvyagin, 2010; Lan et al., 2010). The thermoelectric efficiency ZT = 2.4 has been reached at T = 300K in the p-type semiconductor $Bi_2Te_3/Sb_2Te_3$ in superlattices with quantum wells (Venkatasubramanian et al, 2001); the estimation indicates that the value $ZT \sim 3.5$ has been received at T = 575 K in nanostructured n-type PbSeTe/PbTe with quantum dots (Harman et al., 2000, 2005). It is possible also to include thermotunnel elements (thermal diodes) to nanostructured thermoelectrics in which there exists the electron tunneling through a narrow vacuum or air gap (Tavkhelidze et al., 2002). The efficiency *ZT=1.7* at the room temperature was received experimentally in thermoelements with cold junctions, consisting from the semiconductor branches of p-type $Bi_{0.5}Sb_{1.5}Te_3$ and n-type $Bi_2Te_{2.9}Se_{0.1}$ (Ghoshal et al., 2002a, 2002b). Let us note also the important results received in a set of papers, for example (Shakouri & Bowers, 1997), which specify in perspectives of the use of the emission nanostructures for creation of effective thermoelectric energy converters and coolers. The nanostructuring gives a new way of improvement of the thermoelectric efficiency because the governance of the sizes of nanostructured elements is a new important parameter for influence on the thermoelectric properties of a material.

Unfortunately the best values of ZT that were specified in nanostructures based on superlattices with quantum wells and quantum dots have not been reproduced in one laboratory of the world. On the other hand fabrication of such superlattices uses very expensive technologies; therefore industrial manufacture of such nanostructures is very problematic from the economical point of view. Good values of the thermoelectric figure of merit in thermotunnel devices and in thermoelements with point contacts also have not been reproduced. Therefore the special interest represents a creation of thermoelectric nanostructures by means of an adaptable to streamlined production and a cheap technique. An example of such technology is fabrication of bulk nanostructured thermoelectric samples by ball milling of initial materials with subsequent hot pressing (Poudel et al., 2008; Bublik et al., 2009; Bulat et al., 2008a, 2008b, 2009, 2011b; Minnich et al., 2009; Lan et al., 2010), spark plasma sintering (Bublik et al., 2010a, 2010b) or extrusion (Vasilevskiy et al., 2010). In Ref. (Poudel et al., 2008) the value ZT = 1.4 have been received at T=100$^0$C and ZT = 1.2 at the room temperature in such bulk nanostructured thermoelectrics fabricated from the solid solutions based on p-type Bi-Sb-Te.

Thus from the set forth above it can be concluded that the following reliable preconditions of the obtaining of the high thermoelectric figure of merit in the nanostructured thermoelectrics take place: (a) the experimental results specify a possibility for the achievement of a high thermoelectric figure of merit in nanostructured thermoelectrics of various types; (b) in particular some experimental results confirm the possibility of the obtaining of high figure of merit in bulk nanostructured semiconductors. Experimental and theoretical results that obtained by the authors during lust few years on investigation of bulk nanocrystalline thermoelectrics based on Bi-Sb-Te solid solution including nanocomposites are summarized, systematized and analyzed in the present chapter.

## 2. Experiment

### 2.1 Fabrication of bulk nanocrystalline thermoelectrics
Two stages should be executed for preparation of bulk nanocrystalline materials. At first a powder from nanoparticles should be fabricated, and then it should be consolidated into a bulk sample. A crystalline thermoelectric material with high thermoelectric efficiency should be chosen as an initial material for the nanopowder preparation. In our case the solid solution based on *p*-type $Bi_xSb_{2-x}Te_3$ was selected as the initial material (Bublik et al., 2009, 2010a; Bulat et al., 2008a, 2008b, 2009a, 2009b, 2011b). It was grown up by zone melting method; and the dimensionless figure of merit $ZT = 1.0$ was measured along the C axis at the room temperature in primary samples. The initial crystalline material was grinded and purifying. The mechanoactivation process (the ball milling) is the most convenient and cheap way for fabrication of a nanopowder. We used the high-speed planetary mill AGO-2U to achieve the further superthin crushing and to prepare the nanopowder. Other types of mills: the Activator 2S, Retsch PM 400 also were applied at different stages of the nanopowder preparation. The processing of the powders fabrication in the mill is made by steel spheres which were collided with acceleration up to 90g. Tightly closed containers of the mill rotate in flowing water that protects a material from a warming up. It is necessary to provide absolute absence of the oxidation of the nanopowder. Therefore all operations were spent in the boxing filled with argon.

The duration of the mechanoactivation processing was varied from 30 min till 2 hours. The diffraction analysis has shown that the main sizes of nanoparticles of the powder are 8-10 nm. The following methods of pressing for fabrication of compact samples from highly active ultradisperse powder have been used (Bublik et al., 2009, 2010a; Bulat et al., 2009a): cold pressing of powders with the subsequent sintering in inert gas; sintering in graphite compression moulds; sintering in steel compression moulds (at more high pressure in comparison with the previous variant). Hot pressing of the nanopowder was made under the pressure in the range from 35 MPa to 3.3 GPa in the range of temperatures from 250 to 490 °C. To prevent the oxidation of nanoparticles all basic operations are made in the atmosphere of argon.   As a result, series of compact *p*-type $Bi_xSb_{2-x}Te_3$ samples were produced. The method of spark plasma sintering (SPS) with the equipment SPS–511S for preparation of bulk nanostructure was also used (Bublik et al., 2010b, 2010c).

## 2.2 Methods of experimental investigation

For investigation of thermoelectric properties it is necessary to know values of four material parameters: Seebeck coefficient (thermoelectric power), electric conductivity, heat conductivity and thermoelectric figure of merit. The heat conductivity measurement of small samples is the most difficult because all traditional techniques of direct measurement are based on passing of a calibrated thermal flow through a sample; but a thermal flow measurement with sufficient accuracy and consideration of all losses on small samples is very complicated. Therefore we had been used the Harman method (Bublik et al., 2009, 2010a) which allows to fulfill the measurement of thermoelectric figure of merit Z directly by measurement of only electric parameters, not mentioning about thermal flows. Besides, the technique allows receiving in the same cycle of the measurement the values of Seebeck coefficient and the electric conductivity also. Then the value of heat conductivity can be calculated using the known value of Z. A mathematical model for calculations of thermoelectric parameters on the Harman method measurement has been developed, and processing of results of the measurement was carried out under specially developed soft.

For determining of the speed of longitudinal acoustic waves and for subsequent calculation of modules of elasticity and the modulus of dilatation the modified echo-pulse method with application of focusing system of an acoustic microscope has been used. The mode of ultrashort probing impulses has been utilized; it has given the opportunity to register separate signals caused by the reflexion of the impulse from walls of a sample. The microstructure of samples was investigated at metallographic sections made on grinding-and-polishing machine "Struers". Microhardness was measured on microhardness gauge PMT-3M by the method of cave-in of diamond tips. The microscope Olympus BX51 was used for the metallographic analysis.

The working capacity and reliability of thermoelectric devises are substantially caused by their strength characteristics. The strength at the extension occupies a special place among them. However for investigated materials the method of direct test for the extension is the extremely inexact for some reasons. Therefore the method of diametrical compression of disk or cylindrical samples was used; the advantage of the method consists that the extension pressure destruction begins inside a sample instead of its surface. Determination of the density of samples was made by the method of hydrostatic weighing. Laboratory analytical electronic scales "KERN", model 770-60 were used. The option "Sartorius" was applied for determination of the weight of a solid in a liquid.

The X-ray diffractometer methods were used for investigation of structure of nanopowder and bulk samples. The phase analysis was carried out by the method of X-rays diffraction with the diffractometer Bruker D8, equipped by the scintillation detector Bruker. The lattice constant of a solid solution of a thermoelectric material was determined by shooting of a diffractograms in the standard symmetric scheme of reflexion. A composition of the solid solution on the basis of measurements of the lattice constant was estimated. Values of nanograins were estimated by sizes of coherent dispersion areas (CDA) determined by the method of X-ray diffractometry on broadening of diffraction maxima. Calculation of CDA and estimation of microdeformation were spent by means of Outset program. The received values of CDA size were compared with the data received by a method of high resolution transmission electron microscopy (HRTEM). The structure of a sample was analyzed by means of creation of return polar figures. They were carried out by shooting of diffractograms in the standard symmetric scheme of reflexion. The following microscope equipment was used:  the scanning electron microscope JSM-6480LV with the option for the energy-dispersive analysis INCA DRY; transmission electron microscope JEM 2100 with ultrahigh resolution and X-ray photoelectron spectroscope.

## 2.3  Structure and mechanical properties of nanopowder

Different types of nanopowder from $Bi_xSb_{2-x}Te_3$ solid solutions with different value of x were prepared with the following duration of the mechanoactivation process (the ball milling): 15 min, 30 min, 60 min and 120 min. For each type of powder the X-ray diffractograms, a distribution of CDA size and HRTEM images were received. A typical diffractogram of $Bi_{0,5}Sb_{1,5}Te_3$ powder is presented in (Bulat et al., 2009b). Examples of distribution of CDA size of the nanopowder prepared from $Bi_{0,5}Sb_{1,5}Te_3$ during 60 min ball milling and the correspondent HRTEM image are shown in Fig.1.

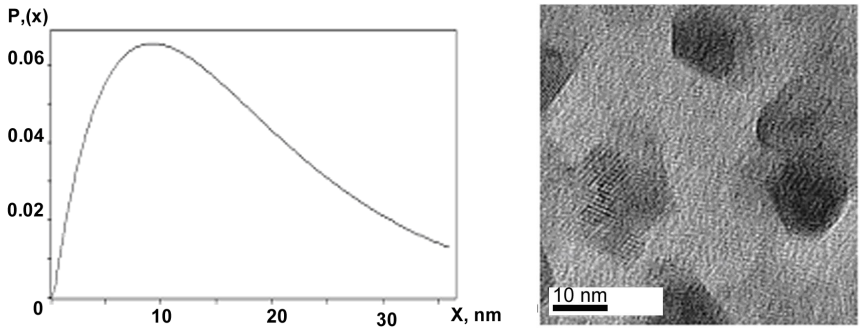

Fig.1. Distribution of CDA size and TEM image of nanoparticles for $Bi_{0,5}Sb_{1,5}Te_3$ (60 min ball milling)

The nanopowder is the single-phase solid solution of $Bi_{0,5}Sb_{1,5}Te_3$ at each duration of the mechanoactivation process. Microdeformations of nanoparticles did not reveal. The electron microscopic data on the average size of nanoparticles confirm the calculation of CDA size determined by broadening of the diffraction maxima. In particular for 2 hours of ball milling the average value of nanoparticles was 8.5 nm and the greatest size was 35 nm. The distribution of sizes is homogeneous enough. An insignificant increase of the average size of CDA in comparison with the powder passed processing during 60 min is observed. A monotonous reduction of the average size of CDA was observed according to increase of the duration of the ball milling processing from 15 min to 60 min, and the size distribution of nanoparticles became more homogeneous. However the further increase of duration of milling leads any more to a reduction but to an increase of the average CDA size. Microstrains have been found out in the powder after 120 min ball milling. The mean-square microstrains are equal to 0.144%. So small particles cannot contain a dislocation therefore the presence of microstrains in the powder can be connected with heterogeneity of a structure of the solid solution, arising at long processing.

The lattice constants of the solid solution for all duration of processing are: a=0.4284 nm and c=3.0440 nm.

## 2.4 Structure and mechanical properties of bulk nanocrystalline samples

Taking into account set forth above the 60 min duration of the milling have been chosen for prepare of bulk samples. Therefore the average size of nanoparticles in the starting powder from $Bi_{0,5}Sb_{1,5}Te_3$ was equal to 8 – 10 nm. The cold and hot pressing and SPS method were applied for preparation of the bulk nanostructures.

The cold pressure under 1.5 GPa was carried out within 60 min (without sintering). The correspondent diffractogram shows that the sample is single-phase one, and it does not contain an exudation of moisture of another phase. The diffraction peaks belong to threefold solid solution of $Bi_{0,5}Sb_{1,5}Te_3$ with the lattice constants: a=0.4284 nm and c=3.0440 nm. The diffraction streaks remained strongly blurring as well as in a powder after the milling. The distribution of CDA size after the cold pressure illustrates Fig.2. The average size of CDA after the cold pressure is equal to 12 nm. As follows from Fig.2 at the cold pressing CDA size does not increase practically, and also uniformity of the nanoparticles size distribution increases. Microdeformations did not reveal.

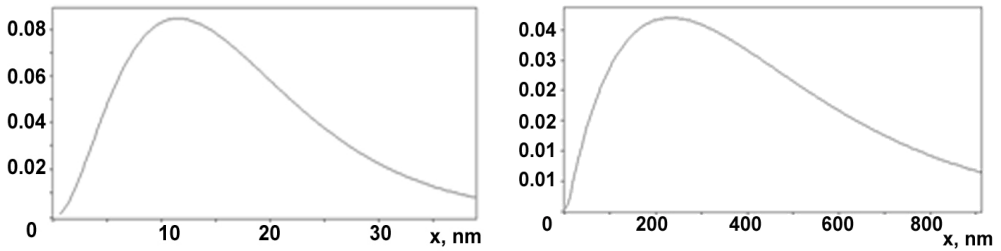

Fig.2 Distribution of CDA size for bulk $Bi_{0.5}Sb_{1.5}Te_3$ after cold pressure; (a) – without sintering, (b) – after sintering

The investigation has shown that the sintering after cold pressing at temperatures ~ $300^0C$ and above leads to two effects: (a) to occurrence of the second phase (tellurium) and (b) to origination of microdeformations. For example after the cold pressing a sample was sintered at the temperature $350^0C$ within 25 min in argon atmosphere. It contains two phases $Bi_{0.5}Sb_{1.5}Te_3$ and tellurium. The lattice constants of the solid solution a=0.4296 nm and c=3.0447 nm are increased in comparison with the lattice constants in the initial sample before sintering. Root-mean-square microdeformation was equal to 0,086 %. Increase of the lattice constants of the solid solution as well as occurrence of microdeformations are apparently results of the excretion of tellurium from the solid solution. Thus as it was marked above microdeformations can be caused by heterogeneity of the solid solution arising at the raised temperature.

Fig.2 shows also the distribution of CDA size for the cold pressed bulk sample of $Bi_{0.5}Sb_{1.5}Te_3$ with the subsequent sintering. The average size of CDA is equal to 230 nm and the maximum size – 900 nm. Character of the size distribution in comparison with the cold pressing without the sintering has not changed, but the curve was displaced towards the big size. The increase of the CDA size testifies that there have passed processes of recrystallization during the sintering at 350 $^0C$.

The SEM images of the sample received by cold pressing are characterized by high porosity (Fig.3). Pores are dark formations as they do not reflect electrons. The size of pores reaches 5 μm. It is visible at a big resolution that pores are not spherical; and coagulation of small pores takes place, therefore a facet is formed. The pores are faceted as a result of the diffusion processes during the sintering.

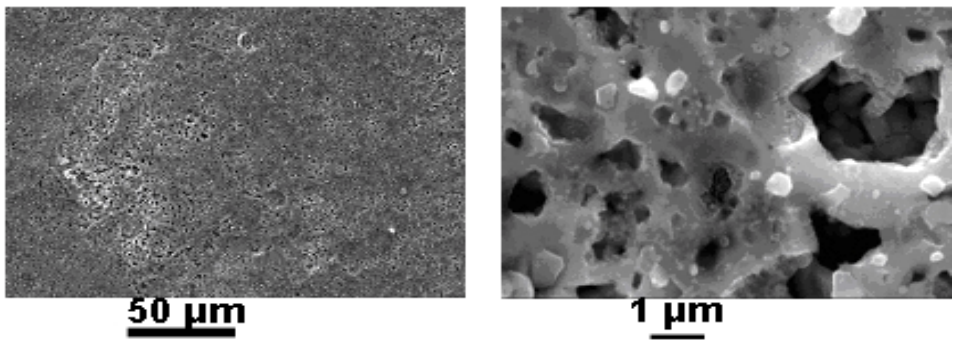

Fig.3. The SEM images of the bulk $Bi_{0,5}Sb_{1,5}Te_3$ (cold pressing with sintering). The resolution is 500 and 10000 accordingly

The CDA size and structure of samples fabricated by the hot pressing are defined by three factors: a temperature of pressing, duration of stand-up under the loading and a value of pressure. But in any case the hot pressing as like as the cold pressing with the subsequent sintering leads to an occurrence of the second phase and to microdeformations. The increase of the CDA size due to the processes of recrystallization also takes part at the hot pressing. Fig.4 shows a typical distribution of CDA size for the hot pressed bulk sample; it was pressed during 20 min at 0.2 GPa and 289 $^0$C. The sample contains two phases: $Bi_{0,5}Sb_{1,5}Te_3$ and tellurium. The lattice constants are increased in comparison with the lattice constants in the initial nanopowder. Root-mean-square microdeformation was equal to 0,055 %, the average CDA size ~ 85 nm, the biggest CDA has the size ~ 300 nm.

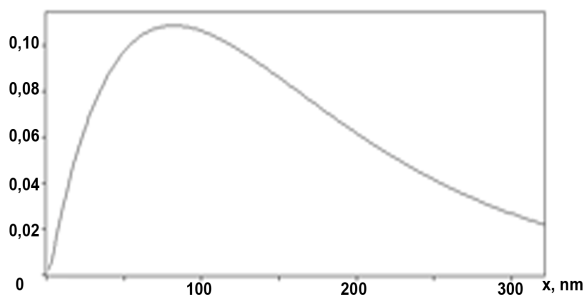

Fig.4. Distribution of CDA sizes of the bulk $Bi_{0,5}Sb_{1,5}Te_3$ (hot pressing)

The SEM images of the surfaces of this sample are shown at Fig.5. A relief specified to different speed of dissolution arises after chemical polishing. The observable elements of structure: consertal formation, micropores, cracks, allocation of a second phase, microdeformations of CDA, connected with a method of hot pressing, are the factors reducing thermoelectric properties of the sample, first of all, the electric conductivity.

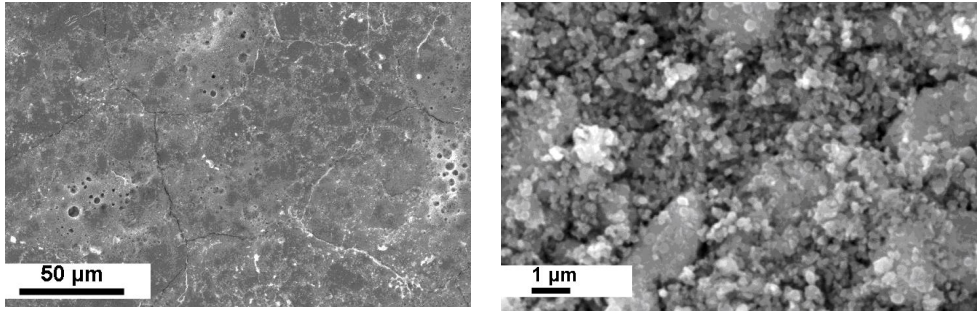

Fig.5. The SEM images of the sample of bulk $Bi_{0,5}Sb_{1,5}Te_3$ (hot pressing). The resolution is 500 and 10000 accordantly

The samples for SPS were prepared by the cold pressure from the nanopowder at the room temperature. Then the SPS processing in a graphite press mould was made by passing of a pulse electric current under the pressure 50 MPa and the temperature 250-400 $^0$C up to achievement of the hundred-percent density (from theoretical value for the given material). The correspondent SEM immerges can be seen in Fig.6.

At the temperatures of sintering 350 $^0$C and 400 $^0$C the grains grow and facet; that testifies about the active process of recrystallization. For samples that were sintered at a lower temperature the finely divided structure is typical, fragments of the fracture surface are not faceted, i.e. the grains have not recrystallized yet.

The average density of the samples fabricated from $Bi_{0,5}Sb_{1,5}Te$ solid solution by cold and hot pressing (plus 4 mass % of Te) are presented in table 1; the accuracy is ±0.02 g/cm$^3$. We see that increase of the temperature and increase of the pressure lead to gain of the density almost to the density of initial samples. However according to the ultrasonic microscopy microdefects in the form of separate cracks are found out even in fabricated at the high temperature and pressure samples. Such defects can lead to the decrease of the strength of nanostructured samples and can lead to reduce of the density. The values of elastic modules are presented in table 2.

| Initial samples | Cold pressing at 1.5 GPa without sintering | Cold pressing at 1.5 GPa with subsequent sintering at 350ºC | Hot pressing at 35 MPa and 470ºC | Hot pressing at 250 MPa and 490ºC |
|---|---|---|---|---|
| 6.71 | 5.02 | 5.60 | 6.62 | 6.69 |
| 6.45 | 5.00 | 5.67 | 6.41 | 6.64 |
| 6.69 | 5.12 | 5.86 | 6.48 | 6.70 |

Table 1. Density of samples fabricated under different modes, g/cm$^3$

| Mode of sample fabrication | Elastic modules | | | | | | |
|---|---|---|---|---|---|---|---|
| | $V_L$ | $V_T$ | $\rho V_L^2$ | $\rho V_T^2$ | B | E | $\Sigma$ |
| Initial sample | 3,26 | 1,77 | 71,84 | 21,1 | 43,7 | 54,5 | 0,292 |
| Cold pressing at 1.5 GPa with subsequent sintering at 350ºC | 2,94 | 1,75 | 58,17 | 20,61 | 30,7 | 50,5 | 0,226 |
| Hot pressing at 350 MPa and 470ºC | 3,46 | 2,16 | 80,09 | 31,21 | 38,48 | 73,65 73,71 | 0,181 |
| Hot pressing at 250 MPa and 490ºC | 2,64 | 1,55 | 41,96 | 14,46 | 22,68 | 66,19 | 0,237 |
| Hot pressing at 250 MPa and 490ºC | 2,72 | 1,59 | 44,54 | 15,22 | 24,25 | 69,36 | 0,240 |

Table 2. Elastic modules of samples fabricated under different modes, GPa

The change of the elasticity module E (Young's module) can be caused by changes of the concentration and sizes of defects in the type of micro- and sub-microcracks, formed at consolidation of nanostructured materials. It proves to be true according to the direct supervision of the samples structure (Fig.6).

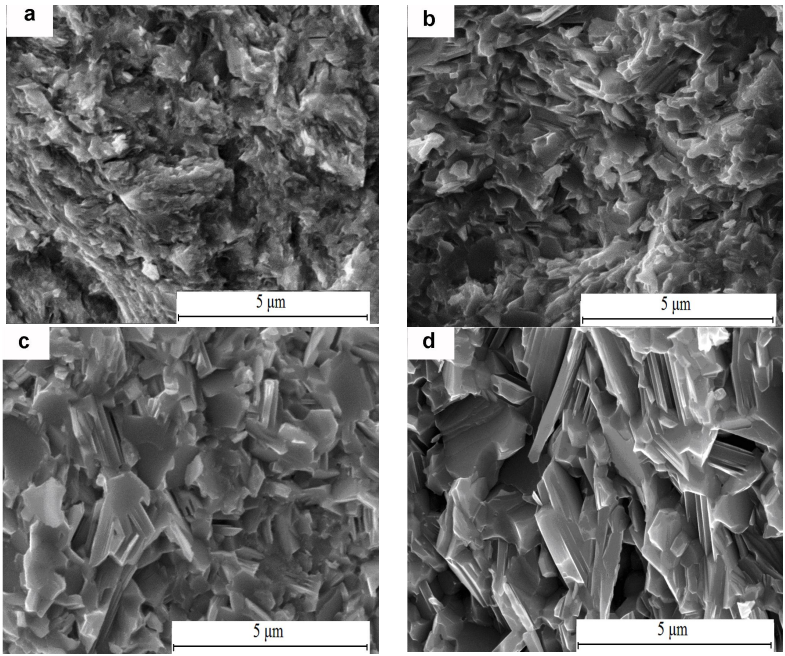

Fig.6. SEM fractographs of sintered $Bi_{0,4}Sb_{1,6}Te_3$ samples at the pressure 50 MPa. SPS temperatures: a) 240 °C, b) 300 °C, c) 350°C, d) 400 °C.

The received by SPS method samples were strong mechanically at all temperatures of sintering. Pores were absent. Results of the samples testing on diametrical compression are presented in Table 3.

| Samples cut out from an ingot | Samples after cold pressing and sintering | Samples after hot pressing at 350 MPa and 470°C |
|---|---|---|
| 6.8 | 20.3 | 27.3 |
| 11.4 | 15.0 | 31.0 |
| 4.5 | 22.6 | 29.5 |
| | 19.7 | |

Table 3. Compressive at diametrical strength (MPa)

## 2.5 Structure of bulk nanocomposites

Thermoelectric properties of materials with nanocrystalline structure should depend essentially on the size of nanograins (CDA) in a bulk sample. In turn the size of grain is defined by a number of factors: a temperature of the sintering or the hot pressing, duration of the hot pressing, value of the pressure, composition of materials, including presence of nano-adding of a second phase in a composite material. It is possible to ascertain that fabrication of initial nanopowder is less complex technological problem than maintain of the bulk nanostructure during the hot pressing that is caused by growth of initial nanoparticles due to recrystallization.

We see from Sec.2.4 that that the samples sintered after the cold as like as the hot pressure leads to increase of CDA size (or nanograins' size). In general nanostructured material is nonequilibrium by its nature, therefore thermal influences (at a manufacturing or an operation) are usually accompanied with the recrystallization of a compact material and a degradation of its properties. A possible way to reduce the average size of nanograins can be an inclusion of nanoparticles from another chemical composition, it means fabrication of nanocomposites. To investigate the relative change of nanograin size we added another nanoparticle-phase to the same solid solution matrix. They were added before the mechanical activation process. Three types of the extra nanoparticles were used for fabrication of nanocomposites: (a) $MoS_2$ with a laminated structure; (b) fullerene $C_{60}$, and (b) thermally expanded graphite (TEG). Values of nanograins was estimated by sizes of coherent dispersion areas (CDA) determined by the method of X-ray diffractometry on broadening of diffraction maxima. The received values of CDA sizes were compared to the data, obtained by the method of high resolution transmission electron microscopy (HRTEM). Both methods have shown a good consent of results at least at the size of grains up to several tens in nanometer. Larger grains also will consist from CDA with various crystallographic orientations which still influence on physical properties.

The content of $MoS_2$ was varied from 0.1 to 0.4 mass %. Only the peaks belonging to the triple solid solution $Bi_{0,4}Sb_{1,6}Te_3$ can be seen in the X-ray diffractogram of such a nanocomposite. The lattice constant of the nanocomposite does not change. Such a situation is repeated regardless of the pressing temperature. TEM study shows that $MoS_2$ nanoparticles are situated at the grain boundary, and do not dissolve in the matrix (Fig.7).

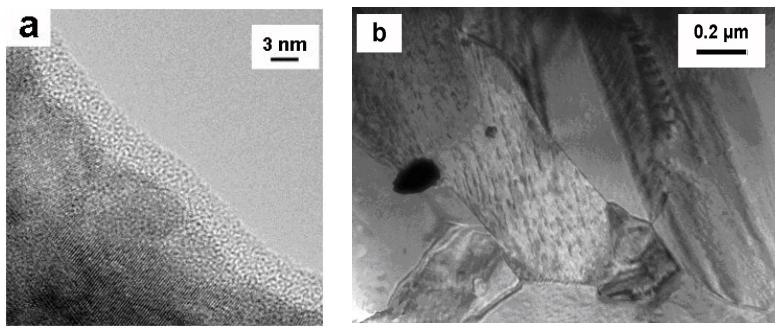

Fig.7. (a) - HRTEM image of $Bi_{0.4}Sb_{1.6}Te_3$ nanoparticle covered by levels of $MoS_2$; (b) - image of the $Bi_{0.4}Sb_{1.6}Te_3$ sample fabricated at 350 MPa and $350^0C$ with 0.1 mass % of $MoS_2$

The $MoS_2$ particles have sizes ~ 20nm, and they have a crystalline structure. The introduction of $MoS_2$ greatly reduces the average size of nanograins and makes their size distribution more uniform. The maximum size of the nanograins decreases from 180 nm (in the solid solution without additives) to 80 nm (at a content of 0.1 mass % of $MoS_2$). The increase of the contents of $MoS_2$ up to 0.4 mass % leads to a future reduction of the average size of the nanograins and also leads to a more uniform size distribution. The toughness of the sample of the same composition was not less than 150 MPa. The addition of 0.1 mass % of $MoS_2$ brought about an increase of the toughness of 20-30 %.

The fullerene $C_{60}$ (1,5 mass %) or the thermally expanded graphite (TEG) (0.1 mass %) were added to the micropowder from initial crystalline material $Bi_{0.5}Sb_{1.5}Te_3$ of p-type. Then the mechanoactivation processing was made at different temperatures under pressure 350 MPa during 20 min in the argon atmosphere. The nanopowder received without carbon additives represented 100÷300 nm units consisting in turn from nanoparticles. The average CDA size was 8÷10 nm.

The mechanoactivation of $Bi_{0.5}Sb_{1.5}Te_3$ samples in the presence of TEG was accompanied by a stratification of the graphite and a formation of flakes with the size of few nanometers; layers from the graphite flakes cover the semiconductor nanoparticles. Fig.8 shows the size and the configuration of carbon layers. The received layered covers on the semiconductor nanoparticles had as the ordered (similarly as layer of graphite on a surface) and the disorder structure. Let us notice that formation in the same process of the ordered and the disordered carbon covers is undesirable as they have different type of conductivity. This factor can cause a bad reproducibility of the properties of thermoelectric nanocomposites. Unlike TEG the fullerenes possess strongly pronounced electrophilic properties; therefore it would be interesting to track a combination of this form of carbon with the semiconductors' nanoparticles. The state of the interface «semiconductor – $C_{60}$ – semiconductor» can make an essential impact on the transport properties at the expense of change of an electronic condition in thin layers of nanoparticles without chemical doping (Bulat et al., (2006). Nanoparticles from the semiconductor covered by layers with disorder structure from $C_{60}$ molecules have been received by the mechanoactivated processing of $Bi_{0.5}Sb_{1.5}Te_3$ together with the fullerenes. The typical structure of such particle is shown at Fig. 9.

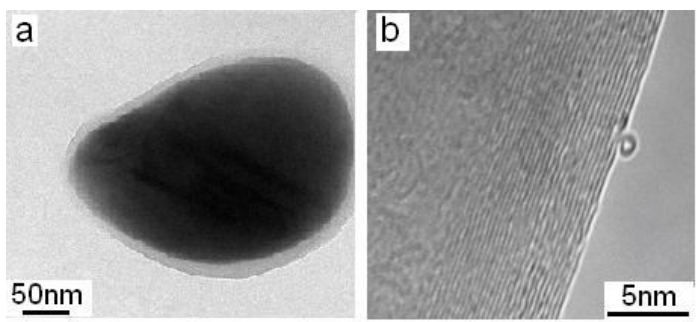

Fig.8. Carbon covers from mechanoactivated TEG on the surface of $Bi_{0.5}Sb_{1.5}Te_3$ nanoparticles: (a) – ordered and (b) - disordered structure

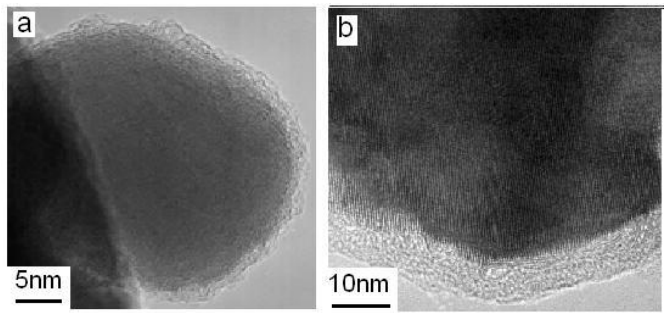

Fig. 9. HRTEM images of a semiconductor particle in a cover from molecules $C_{60}$: (a) - $Bi_{0.5}Sb_{1.5}Te_3$; (b) - $Bi_{0.4}Sb_{1.6}Te_3$

It has been determined, that at mechanoactivation processing of $Bi_{0.5}Sb_{1.5}Te_3$ solid solution the additive of nanocarbon do not influence to the average size of CDA; in all cases it was 8÷10 nm. However the application of nanocarbon has allowed to reduce essentially disorder of CDA size, and to reduce in 1.5÷2 times a share of concerning large (more then 30 nm) particles. Apparently the received carbon covers effectively break the recrystallization of nanoparticles: the CDA size have decreased in 1.7÷1.9 times at the temperature 400÷450°C.

The samples cut out from an initial ingot have shown a considerable disorder of the strength $\sigma_p=0.5\div2.5$ MPa. The nanostructures samples from $Bi_{0.5}Sb_{1.5}Te_3$ had the strength $\sigma_p=18.5\div20$ MPa, and for $Bi_{0.5}Sb_{1.5}Te_3$ samples with TEG and with $C_{60}$ the value of strength 26.3 and 31.0 MPa accordingly have been received.

Fig. 10 shows the consolidated data of the influence of different factors on the temperature dependencies of the average nanograins size for p-type nanocrystalline samples that were fabricated under different pressing mode (Bublik et al., 2009; Bulat et al., 2010b). We see that the main factors allowing slow growth of nanograins as a result of recrystallization are the reduction of the temperature and of the duration of the process, the increase of pressure, as well as the addition of small amount of additives (like $MoS_2$, TEG or fullerenes). In the case of additives the accidental particles in a nanocomposite settle down on borders of particles of the basic solid solution creating the structure like "core – cover". Let us underline that the CDA size coincides with the size of grains revealed on the SEM image of the break of surface in a compact sample at the sizes of grains to several tens nm. In the larger grains ~ 1–2 μm CDA are a part of the internal structure of a grain.

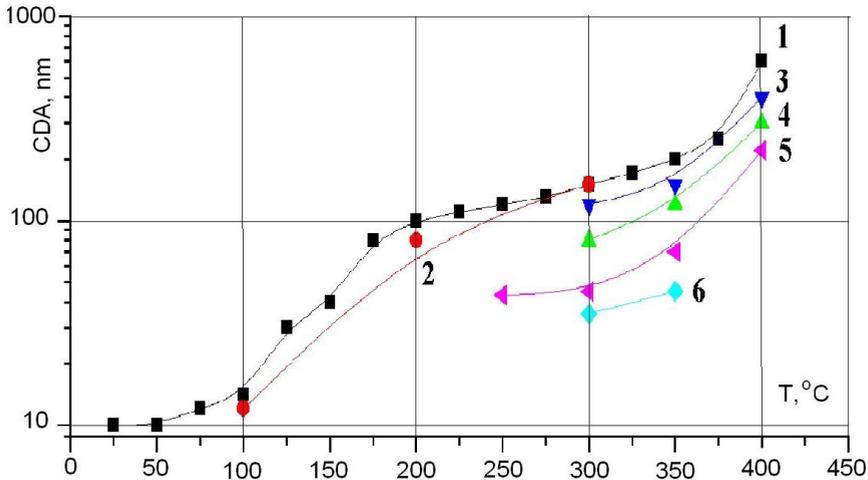

Fig.10. Temperature dependence of the average size of nanograins of samples fabricated under different modes of pressing. 1 – in situ heating of  $Bi_{0,4}Sb_{1,6}Te_3$  samples pressed at $25^0$C and 1.5GPa in the thermocamera of the diffractometer; 2 – vacuum annealing at different temperatures of  $Bi_{0,4}Sb_{1,6}Te_3$  samples pressed at $25^0$C and 1.5GPa; 3 –  $Bi_{0,4}Sb_{1,6}Te_3$  samples hot pressed at 35MPa; 4 –  $Bi_{0,4}Sb_{1,6}Te_3$  samples hot pressed at 350MPa; 5 –  $Bi_{0,4}Sb_{1,6}Te_3$  samples plus 0.1 mass% $MoS_2$ pressed at 350MPa; 6 –  $Bi_{0,4}Sb_{1,6}Te_3$  samples plus 0.4 mass % $MoS_2$ pressed at 350MPa

As properties of a material to a great extent depend on its structure in micro- and nanoscale, the comparative analysis of the structure of the bulk samples received by SPS methods and traditional hot pressing has been carried out; these results are presented at Fig. 11, 12. The analysis of the received results shows that unlike the method of hot pressing the SPS method allows to receive at rather low pressure 50 MPa mechanically strong well sintered nanostructured materials. It does not contain pores even at temperatures more low then 300˚C. The explanation of this result is that in SPS process the high density of allocated energy in contact zones between the powder particles causes the very strong local warming up (up to fusion of a grain blanket) whereas the basic volume of a the material remain at lower temperature. The CDA size for both methods of consolidating up to temperatures 350˚C are comparable, whereas at sintering temperatures 400˚C and above (that in practice corresponds to temperatures of hot pressing) the CDA size in SPS method increases much less.

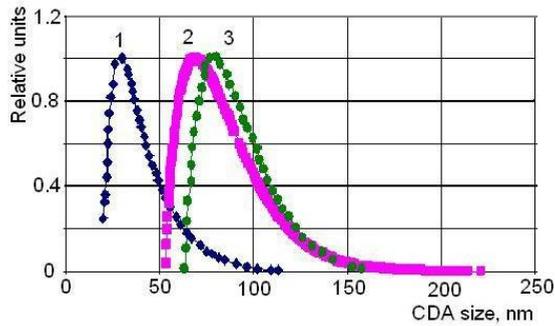

Fig. 11. Relative quantitative portion of different CDA size in nanostructured sintered $Bi_{0,4}Sb_{1,6}Te_3$. SPS temperatures: 1 - 240 °C; 2 - 300 °C; 3 - 350 °C; pressure 50 MPa

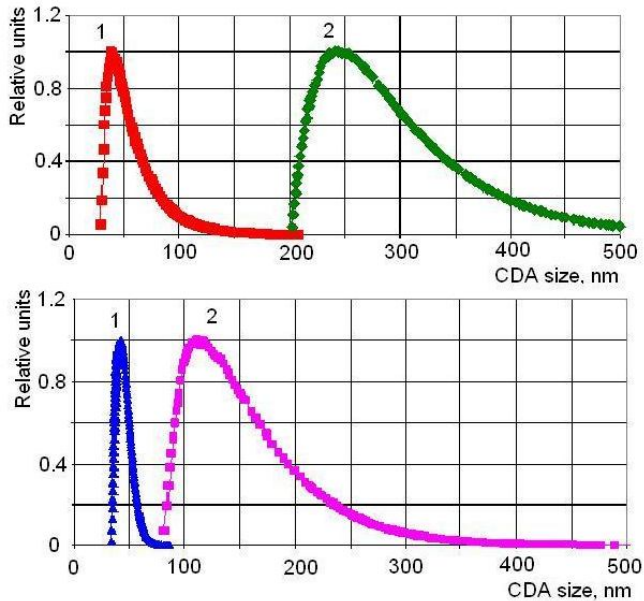

Fig. 12. Relative quantitative portion of different CDA size in hot pressed nanostructured $Bi_{0,4}Sb_{1,6}Te_3$ (350 MPa; 1-300°C &, 2- 400°C). a) $Bi_{0,4}Sb_{1,6}Te_3$, b) $Bi_{0,4}Sb_{1,6}Te_3$+0,1 mass % $MoS_2$

## 2.6 Thermoelectric properties of bulk nanostructures and nanocomposites

The main transport properties of nanocrystalline materials fabricated under different conditions were investigated. Dependences of the transport properties on average nanograins size were also analyzed.

The temperature dependences of the thermoelectric parameters of typical hot pressed nanostructured p-$Bi_{0,3}Sb_{1,7}Te_3$ sample was published and discussed in Ref. (Bulat et al., 2010b). The correspondent maximum value ZT=1.12 takes place at the temperatures ~ 350÷375 K. The same maximum efficiency ZT=1.1 in the same nanostructured material at the same temperature was measured in Ref. (Vasilevskiy et al., 2010) but the extrusion instead of hot pressing for consolidation of samples was used here.

Let as consider more in detail our investigation of thermoelectric properties of samples fabricated by the SPS method. Thermoelectric properties were studied depending on sintering temperature on samples of p-type $Bi_{0.5}Sb_{1.5}Te_3$ and $Bi_{0.4}Sb_{1.6}Te_3$. All samples have been received by SPS method at pressure 50 MPa, temperature from 250 to 500 $^0$C, the duration of sintering was 5 min (for $Bi_{0.4}Sb_{1.6}Te_3$) and 20 min (for $Bi_{0.5}Sb_{1.5}Te_3$). The samples sintered at 300 $^0$C have the maximum value of thermoelectric power. A small distinction exists between values of thermoelectric power for samples of various compositions. The electric conductivity increases with the raise of sintering temperature. The similar dependence is observed also for the heat conductivity; increase of the heat conductivity with rise of the temperature sintering is caused by increase in the electronic heat conductivity which is proportional to the electric conductivity.

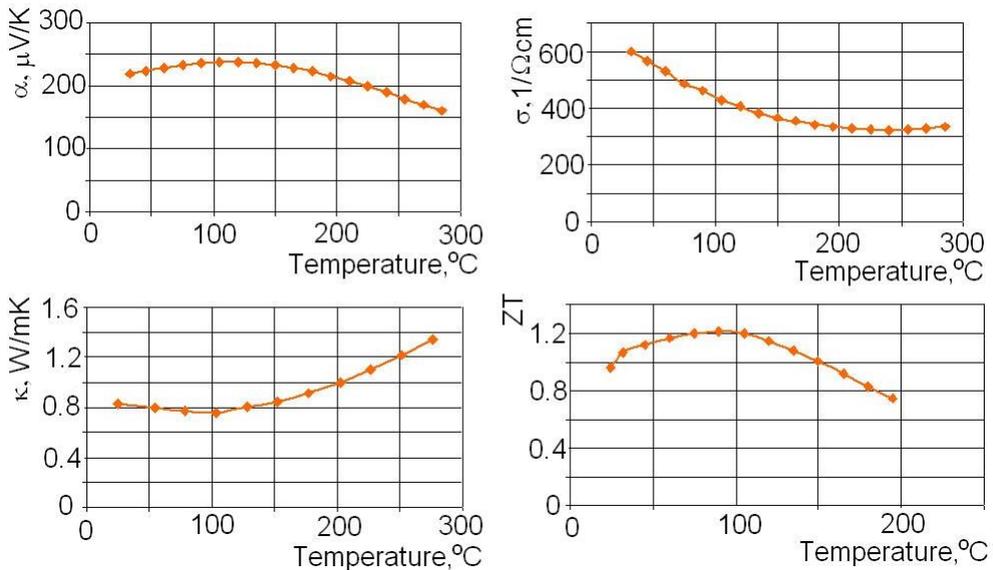

Fig. 13. Thermoelectric properties of sintered bulk nanostructured materials $Bi_{0.4}Sb_{1.6}Te_3$ as a function of measurement temperature: a) electrical conductivity, b) Seebeck coefficient, c) thermal conductivity, d) figure of merit, ZT.

It follows from experimental results that the samples received by SPS method at sintering temperature 450 $^0$C have the greatest value of the efficiency ZT. The samples fabricated from the 0.5 μm powder have lower efficiency ZT in comparison with the nanostructured material. It was established that the pressure 50 MPa is the optimum one for obtain the high thermoelectric efficiency. Samples of $Bi_{0.4}Sb_{1.6}Te_3$ composition obtained at the sintering temperatures ~ 350 $^0$C have higher ZT than $Bi_{0.5}Sb_{1.5}Te_3$ samples. The temperature dependence of thermoelectric parameters in nanostructured samples $Bi_{0.4}Sb_{1.6}Te_3$ received at sintering temperature 400 $^0C$ and pressure 50 MPa is presented at Fig. 13. Peak efficiency is reached at 90 $^0C$ and makes ZT=1.22.

Some dependences of measured thermoelectric coefficients in bulk nanostructured materials on the grain size for solid solution $Bi_xSb_{2-x}Te_3$ (Bulat et al., 2010c, 2011a) will be presented in Sec. 3.2 and 3.3.

## 3. Theory

Three mechanisms that can improve the thermoelectric efficiency are studied theoretically and compared with the experiment in Sec.3.

### 3.1 Electron tunneling

One of the possible mechanisms of electric transport in nanostructured materials is the tunneling of charge carriers through the intergrain barriers. This effect is similar to the thermionic or field emission thought the vacuum gap. The studies of the thermionic emission applied to the field of energy conversion began in the 1960th (Anselm, 1951). Though the efficiency of thermionic generators can reach 20% their working temperatures are about 1000K because of the large values of work function in metals and semiconductors.

To use thermionic devices at lower temperatures one should decrease the work function, e.g. by applying high electric field (Fleming & Henderson, 1940; Murphy & Good, 1956), by using special cathode coatings that can decrease the work function down to 0.8eV (Sommer, 1980) or by utilizing the tunneling effect through the thin vacuum gap (Hishinuma et al., 2001; Tavkhelidze et al., 2002). As was shown by Mahan (Mahan, 1994) to use thermionic devices for refrigeration at room temperature one needs to decrease the work function down to the values of 0.3-0.4eV that are not available at the present time. But in the case of nanoscale tunneling junction the tunneling probability increases and the noticeable cooling power can be reached even at the work functions of about 0.8eV (Hishinuma et al., 2001). One of the possible cooling applications of such device that consisted of metallic tip over the semiconducting plate was described in (Ghoshal, 2002b). Alternatively Schottky barriers or semiconductor heterostructures can be used instead of vacuum barriers (Mahan & Woods, 1998; Mahan et al., 1998). In such structures the barrier energy height can be as low as 0.1eV but the phonon thermal conductivity of semiconducting barrier will increase the total thermal conductivity and produce negative influence on the figure of merit.

In this section the bulk nanostructured material that consists of grains separated by tunneling junctions is considered. The influence of the charge carrier tunneling on the thermoelectric figure of merit of such material is theoretically investigated. The shape of nanoparticles is modelled by two truncated cones with the same base (Fig. 14) that allows one to perform calculations in an analytical way (Bulat & Pshenai-Severin, 2010a). The calculations of the heat flow inside the nanoparticles take into account the difference between the electron and phonon temperatures in the limiting case of vacuum gap when phonons cannot tunnel through the barriers.

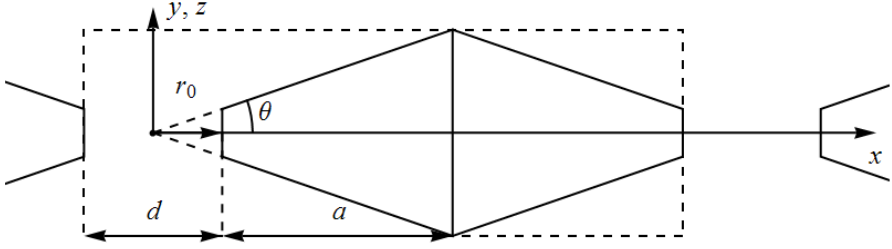

Fig. 14. The cross section of nanoparticle modeled by two truncated cones with the same base. $2a$ is the size of nanoparticle, $2\theta$ is the cone aperture angle, the radius $r_0$ determines the size of truncated part, $d$ is the tunneling junction width. Dashed rectangle represents the single period of the whole structure.

In order to estimate the thermoelectric figure of merit of such material the transport coefficients were calculated. The use of the linear approximation for tunneling coefficients in bulk nanostructured materials seems to be quite reasonable. Indeed if the typical size of samples are several millimeters and the grain size is about 10-40 nm then the voltage drop on single junction is $10^5$ times smaller than on the whole sample. So even at several hundreds of volts bias on the sample the voltage drop on the single junction will be $\Delta V \sim 10^{-3}\,\mathrm{V}$. Similarly the temperature difference on single junction is about $10^{-3}$K at the total temperature difference of 100K. So at the room temperature if the barrier energy height $\varepsilon_b \sim 0.1$eV than $|\,k_0\Delta T\,|$ and $|\,q_0\Delta V\,|$ are much less than both the thermal energy $k_0\,T$ and barrier height $\varepsilon_b$. In this case one can use linear transport coefficient theory for the tunneling junction. The total current density through the tunneling junction is determined by the difference of emission currents from two electrodes (Burstein, 1969)

$$j_x = \sum_{\mathbf{k},\,v_x>0} 2\,(q_0\,v_x)\,D(\varepsilon_x)\left[\,f_0\!\left(\frac{\varepsilon-\mu}{k_0\,T}\right) - f_0\!\left(\frac{\varepsilon-(\mu+q_0\,\Delta V)}{k_0\,(T+\Delta T)}\right)\right]. \qquad (1)$$

In this expression $x$-axis is directed at the right angle to the junction cross section (Fig. 14), $\varepsilon$ is the total energy of electron with the wave vector $\mathbf{k}$, $\mu$ is the chemical potential of the left electrode, $f_0$ is the Fermi-Dirac distribution function, $v_x$ and $\varepsilon_x$ are velocity and kinetic energy of electron corresponding to its motion along $x$ direction and $D(\varepsilon_x)$ is the tunneling probability. In order to obtain linear transport coefficients the expression for the total current density (1) was linearized with respect to small voltage $\Delta V$ and temperature $\Delta T$ differences. Finally the barrier electric conductivity can be obtained as $\sigma_b = -j_x / \Delta V$ (Bulat & Pshenai-Severin, 2010a)

$$\sigma_b = \frac{q_0^2\,m\,k_0\,T}{2\pi^2\,\hbar^3}\int\limits_0^\infty D(\varepsilon_x^*)\,f_0(\varepsilon_x^* - \mu^*)\,d\varepsilon_x^*\,, \qquad (2)$$

where energy $\varepsilon_x^*$ and chemical potential $\mu^*$ with asterisks are measured in $k_0\,T$ units and the effective mass of electron $m$ is assumed to be the same inside nanoparticle and barrier.

The Seebeck coefficient can be obtained from the zero current condition $\alpha_b = -(\Delta V / \Delta T)_{j_x=0}$ and it was expressed as $\alpha_b = \beta_b / \sigma_b$, where (Bulat & Pshenai-Severin, 2010a)

$$\beta_b = \frac{k_0}{q_0} \frac{q_0{}^2 \, m \, k_0 \, T}{2 \, \pi^2 \, \hbar^3} \int\limits_0^\infty D(\varepsilon_x^*) \left[ u \, f_0(u) + \ln\!\left(1 + e^{-u}\right) \right] d\varepsilon_x^* \,, \tag{3}$$

and the following notation was introduced $u = \varepsilon_x^* - \mu^*$.

The expression for electronic heat flow through the junction can be obtained from (1) after replacing $q_0 \, v_x$ with $(\varepsilon - \mu) v_x$. The value of barrier thermal conductivity measured at zero current $\kappa_b$ can be expressed through the thermal conductivity at zero voltage drop $\kappa_{b,\Delta V=0}$ as (Bartkowiak & Mahan, 1999)

$$\kappa_b = \kappa_{b,\Delta V=0} - \alpha_b{}^2 \, \sigma_b \, T \,, \tag{4}$$

where (Bulat & Pshenai-Severin, 2010a)

$$\kappa_{b,\Delta V=0} = \frac{m \, k_0{}^3 \, T^2}{2 \, \pi^2 \, \hbar^3} \int\limits_0^\infty D(\varepsilon_x^*) \left[ u^2 f_0(u) + 2 u \ln\!\left(1 + e^{-u}\right) - 2 \, \mathrm{Li}_2\!\left(- e^{-u}\right) \right] d\varepsilon_x^* \,, \tag{5}$$

and the dilogarithm function is denoted as $\mathrm{Li}_2(x)$. It is worth to note that the barrier electrical and thermal conductivities are determined with respect to voltage and temperature difference instead of their gradients as in the bulk case. Hence for the case of comparison with the bulk values it is more convenient to use the values of $\sigma_b \, d$ and $\kappa_b \, d$.

In the present calculations the intergrain barrier shape was assumed to be rectangular. In linear operating region the change of the tunneling barrier shape under applied field can be neglected. So the well-known expression for tunneling probability of rectangular barrier was used

$$D(\varepsilon_x) = \left( 1 + \frac{\left(k_x{}^2 + k_b{}^2\right)^2}{4 \, k_x{}^2 \, k_b{}^2} \sinh^2\!\left(k_b \, d\right) \right)^{-1} \,, \tag{6}$$

where $k_x = \sqrt{2 \, m \, \varepsilon_x} \, / \hbar$ and $k_b = \sqrt{2 \, m (\varepsilon_b - \varepsilon_x)} \, / \hbar$. Note that if $\varepsilon_x > \varepsilon_b$ the wave vector became pure imaginary $k_b = i \, |k_b|$ and hyperbolic sine should be changed to $\sin(|k_b| d) / i$. Very often instead of exact expression for tunneling probability the WKB approximation is used $D_{\mathrm{WKB}}(\varepsilon_x) = \exp(-2 \, k_b \, d)$. In WKB approximation tunneling probability for $\varepsilon_x > \varepsilon_b$ is equal to unity. In the following the values of tunneling transport coefficients calculated using these two approximations will be compared.

Fig. 15-17 show the dependencies of barrier electrical conductivity, Seebeck coefficient and Lorenz number on the size of tunneling junction for different barrier heights 0.4 and 0.8eV. In these estimations the effective mass was equal to $m = 0.7 \, m_0$ that corresponds to the typical hole effective mass in $Bi_2Te_3$ (Goltsman et al., 1972). Doping impurity concentration was equal to $10^{19} cm^{-3}$ for chemical potential close to the band edge. The curves plotted using exact expression for tunneling probability (6) and obtained in WKB approximation illustrate

noticeable difference of two approaches. For metallic electrodes electron energies close to the Fermi level are important and WKB approximation can be used for small tunneling probabilities $D(\varepsilon_x) < 1/e$ as was stated in (Stratton, 1962). For semiconducting electrodes when the charge carrier energies are close to the band edge the preexponential factor can also be important because it approaches zero for small carrier energies. When the tunneling junction width becomes larger these difference decreases because the contribution of small energy carriers is less important.

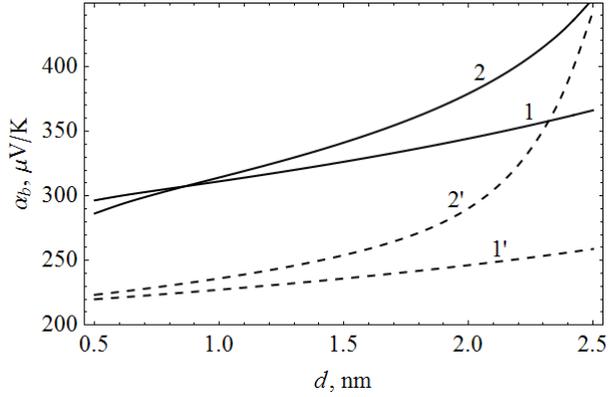

Fig. 15. The dependence of the barrier Seebeck coefficient on the tunneling junction size at room temperature for $\varepsilon_b = 0.8\text{eV}$ (1, 1′) and 0.4eV (2, 2′) calculated using exact expression for tunneling probability (1, 2) and WKB approximation (1′, 2′).

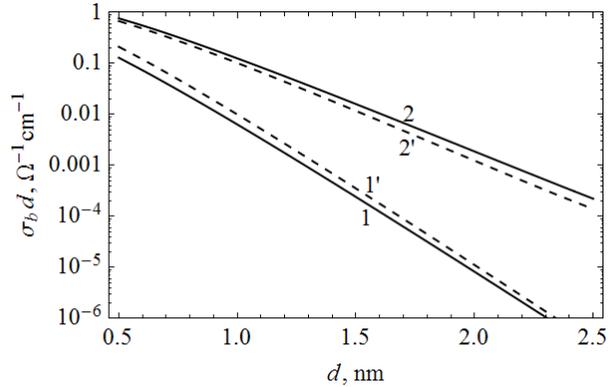

Fig. 16. The dependence of the barrier electrical conductivity on the tunneling junction size (see Fig. 15 for notation).

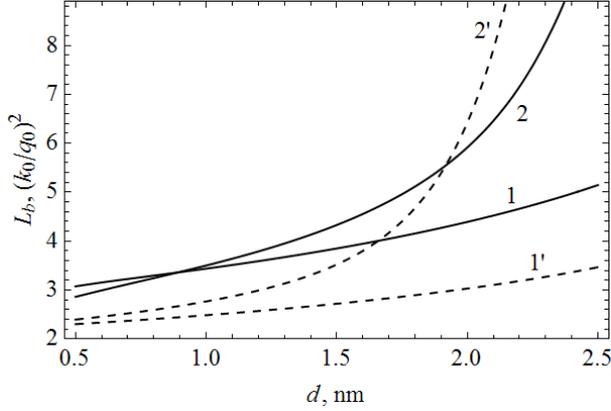

Fig. 17. The dependence of the barrier Lorenz number on the tunneling junction size (see Fig. 2 for notation).

As can be seen from Fig. 15 at the junction thicknesses smaller than 2nm the Seebeck coefficient can reach the values of about 300-350μV/K and slowly varies with the barrier thickness. The electrical conductivity is small (Fig. 16) and decreases exponentially with the barrier thickness. At larger junction thicknesses $d$>2nm the thermionic emission becomes more important than tunneling. Charge carriers with small energies are filtered out of the current that leads to the sharp increase in Seebeck coefficient and Lorenz factor (Fig. 15, 17). The dimensionless figure of merit for single junction is rather high $ZT \approx 3-4$.

In order to calculate effective transport coefficients in the whole structure charge and heat flow inside nanoparticles should be taken into account. As the approximation of zero phonon thermal conductivity of the barrier is considered the heat flow through the junction is only due to charge carriers. Hence the equations for the heat flow should take into account the differences in electron $T_e$ and phonon $T_p$ temperatures (Ghoshal, 2002b; Bartkowiak & Mahan, 1999). The electron-phonon scattering inside nanoparticle leads to the equilibrating of their temperatures on the length scale $l_c$ which is called cooling length. The general solution for conical geometry was obtained in (Ghoshal, 2002b) where the limiting case $l_c << a$ was analyzed. In the considered materials based on bismuth antimony telluride solid solutions the values of cooling length are 66nm for Bi₂Te₂ and 156nm for Sb₂Te₃ (da Silva & Kariany, 2004). So for nanoparticle size of 10-20nm the limit of $l_c >> a$ can be considered. For this case the heat transfer equations were solved for each of two truncated cones representing the nanoparticle in (Bulat & Pshenai-Severin, 2010a). As a result the equations for total resistance and thermal conductance of nanoparticle were obtained (Bulat & Pshenai-Severin, 2010a)

$$R_n = \left(\sigma \gamma \Omega r_0^2\right)^{-1}, \quad K_n = \kappa_e \xi \gamma \Omega r_0^2, \tag{7}$$

where $\gamma = (r_0 + a) / r_0 L_n$, $L_n = 2a$ is the total length of nanoparticle in the $x$-axis direction, $r_0$ determines the size of truncated part (Fig. 14), $\Omega = 4\pi \sin^2 \theta / 2$ is the cone solid angle and

$$\xi = 1 + \frac{\kappa_p}{\kappa_e + \kappa_p} \frac{(a/l_c)^2}{3\gamma L_n} . \tag{8}$$

From these equations it can be seen that in the limit of the small nanoparticle size compared to the cooling length the electrical resistance does not change due to the difference in $T_e$ and $T_p$. The correction to the thermal conductance due to this effect is only second order of magnitude with respect to small parameter $a/l_c$ and $K_n$ is determined mainly by electronic contribution.

It is interesting to note that the transition to the layered geometry can be obtained if $r_0 \to \infty$ and $\theta \to 0$ in such a way that the area $\Omega r_0^2$ is constant. In this limit $\gamma L_n = 1$ and from (7) it is easy to get corresponding equation for the layered system (Anatychuk & Bulat, 2001).

Though in real nanostructured material the size on nanoparticles and their positions are randomly distributed here for estimations of effective transport coefficient the material is modeled as an ordered set of primary cells outlined by dashed lines on the Fig. 14. In this case the total current flow is directed along $x$-axis and effective transport coefficients can be calculated based on equations for layered medium (see, e.g., Snarsky et al., 1997). The effective transport coefficients for the present case were calculated in (Bulat & Pshenai-Severin, 2010a). The thermal conductivity can be obtained as a series connection of barrier and nanoparticle thermal conductivities

$$\kappa_{eff} = \frac{\kappa_b \, \kappa_e \, \xi \gamma}{\kappa_b + \kappa_e \, \xi \gamma} \gamma_t , \tag{9}$$

where geometric factor $\gamma_t = r_0^2 (d + 2a)/(r_0 + a)^2$ was introduced. The effective Seebeck coefficient can be obtain as a sum of Seebeck coefficients of barrier and nanoparticle taking into account corresponding temperature differences on each part

$$\alpha_{eff} = \frac{\alpha_n \, \kappa_b + \alpha_b \, \kappa_e \, \xi \gamma}{\kappa_b + \kappa_e \, \xi \gamma} . \tag{10}$$

In calculations of electrical conductivity the average sample temperature $T_{av}$ is assumed to be constant. But due to Peltier effect the temperatures of neighboring contacts are different. So in the equation for effective electrical conductivity in addition to common expression for series resistance the factor due to Peltier effect induced thermopower should be taken into account

$$\sigma_{eff} = \frac{\sigma_b \, \sigma \gamma \gamma_t}{\sigma_b + \sigma \gamma} \left( 1 + \frac{(\alpha_b - \alpha_n)^2}{\kappa_b + \kappa_e \, \xi \gamma} \frac{\sigma_b \, \sigma \gamma \, T_{av}}{\sigma_b + \sigma \gamma} \right)^{-1} , \tag{11}$$

where $T_{av}$ is the average temperature of sample.

The effective figure of merit of bulk nanostructured material can be calculated using equation (9)-(11) as $Z_{eff} = \alpha_{eff}^2 \, \sigma_{eff} / \kappa_{eff}$. For estimations the typical room temperature parameter for Bi$_2$Te$_3$ from (Goltsman et al., 1972) were used: $\alpha_n$=200 µV/K, σ=830 Ω⁻¹cm⁻¹ and $\kappa_p$=1 W/m K. Though Bi$_2$Te$_3$ is anisotropic material nanocrystals inside the sample are randomly oriented. So for the estimations the values of thermal and electrical conductivities were average over all directions.

On Fig. 18 the dependencies of effective electrical conductivities on the tunneling junction thickness are plotted. It is interesting to note that effective transport coefficients are independent of cone aperture angle $\theta$ because only cross-section areas depend on it and these dependences are canceled out. For larger $r_0$ the electrical conductivity increases due to the increase of the smaller cross-section of the cone. As was noted above in the limit of large $r_0$ the transport coefficients approach the values for layered geometry (compare $i$, $i'$ and $i''$ with $L_i$ on Fig. 18 for $i$=1, 2). For considered parameter range the electrical conductivity of the tunneling junction is much less than the usual values in semiconductors. Hence the effective electrical conductivity is determined mainly by barrier part but it is related to the total period of the structure. For example for layered geometry $\sigma_{eff} \approx (\sigma_b\, d)(L_n + d)/d > \sigma_b\, d$ (compare $L_i$ with $B_i$ on the Fig. 18 for $i$=1, 2). For the case of conical geometry the factor $\gamma_t$ should be taken into account that diminishes $\sigma_{eff}$ for small $r_0$.

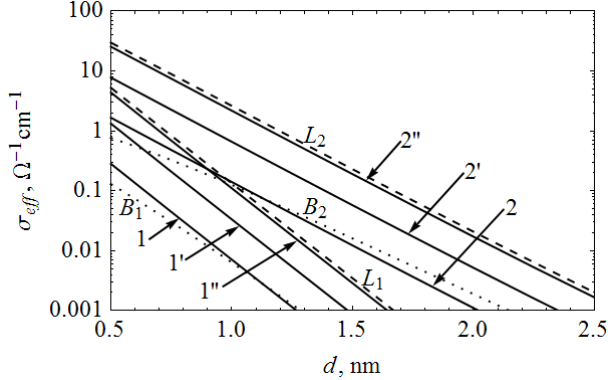

Fig. 18. The dependence of effective electrical conductivity on the tunneling junction thickness for barrier height $\varepsilon_b = 0.8$ (1, 1′, 1′′) and 0.4eV (2, 2′, 2′′) for cone-shaped nanoparticles. The same dependences for single barrier and layer structure are plotted as $B_1$, $B_2$ and $L_1$, $L_2$ correspondingly. The ratio $r_0/a$=0.3 (1,2), 1 (1′, 2′) and 10 (1′′,2′′); 2$a$=20nm.

On Fig. 19 the dependence of effective thermoelectric figure of merit on the tunneling junction thickness is plotted. The estimations showed that in the absence of the phonon thermal conductivity in the barrier for all considered ranges of tunneling junction parameters (see Fig.16, 17) the barrier thermal conductivity is much smaller than the thermal conductivity of nanoparticle. Hence relatively high values of the effective Seebeck coefficient are determined mainly by large $\alpha_b$ (Fig. 15) and the ratio $\sigma_{eff}\,/\,\kappa_{eff}$ in $Z_{eff}$ is determined by the effective Lorenz number that has usual values for small $d$ and begins to increase with the increase of $d$ (see Fig. 17). So in the present case the large values of $Z_{eff}\,T \approx 2.5-4$ are determined by large barrier Seebeck coefficient and the decrease of $Z_{eff}\,T$ for larger $d$ is due to the increase of barrier Lorenz number. Simple estimations of the effect of phonon thermal conductivity of the barrier performed in one-temperature approximation showed that to increase the thermoelectric figure of merit compared to initial semiconducting material the phonon barrier thermal conductivity $\kappa_{b,ph}$ should be about 4 time smaller than the electronic contribution.

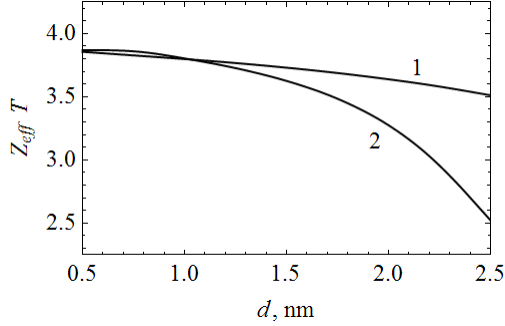

Fig. 19. The dependence of effective thermoelectric figure of merit on the tunneling junction thickness for barrier height $\varepsilon_b = 0.8$ (1) and 0.4eV (2).

To conclude this section it can be said that the thermoelectric figure of merit of the structures with tunneling junctions can be quite large $Z_{eff}\,T \approx 2.5 - 4$ if the barrier phonon thermal conductivity is negligible. These large values are determined by the large values of the barrier Seebeck coefficient and greatly reduced in the presence of $\kappa_{b,ph}$. In addition irregularities in the tunneling junction width or the size of nanoparticles can also lead to the decrease of the figure of merit in real structure. The comparison with the experimental data from Sec.2.6 and Ref. (Poudel et al., 2008; Bulat et al., 2010b; Vasilevskiy et al., 2010) showed that the increase of the figure of merit in these materials is hardly connected with the tunneling effect because of the large difference between measured electrical conductivity (of the order of $1000\,\Omega^{-1}\,cm^{-1}$) and estimated values that are much less than $100\,\Omega^{-1}\,cm^{-1}$.

### 3.2 Boundary scattering

In this section the influence of boundary scattering on the thermal conductivity of bulk nanostructured materials obtained by ball-milling with subsequent hot pressing is considered following (Bulat et al., 2010c). These materials are polycrystalline with small grain sizes in the range from 10 nm to several hundreds on nanometers depending on the temperature of hot pressing. The most common way to estimate the influence of grain boundary scattering on the thermal and electrical conductivities is to include additional scattering mechanism with the mean free path equal to the grain size $L_n$. The theory of this effect applied to thermal conductivity in different polycrystalline solid solutions was described in (Goldsmid et al., 1995) but due to relatively large grain size considered there it was predicted that the effect of boundary scattering on thermal conductivity in bismuth telluride alloys is negligible.

Usually the grain boundary effect is considered to be important for thermal conductivity only at low temperatures when the probability of phonon-phonon scattering decreases. But it is related mainly to pure single crystals (Goldsmid et al., 1995). In solid solutions at high temperatures the contribution of short wavelength phonons to the thermal conductivity is reduced due to the point defect. So in solid solutions the contribution of long wavelength phonons to thermal conductivity is relatively more important than in pure crystals. This contribution can be effectively reduced by introducing boundary scattering.

The estimations performed here are based on Debye model for acoustic phonons with linear spectrum up to Debye frequency $\omega_D$. The following scattering mechanisms are taken into account. In pure single crystals the most important scattering mechanism at room temperature is phonon-phonon umklapp scattering with the relaxation time $\tau_U = A_U / \omega^2$. The thermal conductivity in this case can be written as (Goldsmid et al., 1995)

$$\kappa_0 = \frac{1}{3} c_V \, v_D \, \bar{l}_0 \, , \tag{12}$$

where $c_V = 3 k_0 N_V$ is the heat capacity associated with acoustic modes of the crystal containing $N_V$ primary cells, $v_D$ is the mode averaged Debye speed of sound and $\bar{l}_0$ is the mean free path associated with the umklapp scattering. Knowing the value of thermal conductivity in single crystal $\kappa_0$ the mean free path $\bar{l}_0$ and the constant $A_U$ can be obtained.

When the second component is added forming solid solution the thermal conductivity $\kappa_s$ becomes less than $\kappa_0$ due to the point defect scattering $\tau_P = A_P / \omega^4$. The constant $A_P$ can be deduced from experimental value of $\kappa_s$ assuming that $A_U$ is the same as in initial single crystal. Finally the boundary scattering is described by frequency independent relaxation time $\tau_b = L_n / v_D$. In the simplified treatment (Goldsmid et al., 1995) it was proposed to divide the total range of phonon frequencies into three parts. For each part of the spectrum only the most important relaxation time is considered: $\tau_b$, $\tau_U$ and $\tau_P$ for lower, medium and high frequency parts correspondingly. Then the simple equation for phonon thermal conductivity in polycrystalline material was obtained (Goldsmid et al., 1995)

$$\kappa_{ph} = \kappa_s - \frac{2}{3} \kappa_0 \sqrt{\frac{\bar{l}_0}{3 L_n}} \, . \tag{13}$$

In order to compare the values of $\kappa_{ph}$ with experiment for nanostructured material based on p-Bi$_x$Sb$_{1-x}$Te$_3$ (Bulat et al., 2010c) the hole contribution should be subtracted from experimental values of thermal conductivity. So the proper estimations of electrical conductivity and hole thermal conductivity are necessary. The electrical conductivity in initial solid solution is anisotropic but after ball-milling and hot pressing the samples became isotropic on average. To take the anisotropy into account it was assumed that it is connected mainly with the anisotropy of effective masses and the relaxation time is a scalar. Then using the effective medium theory for average electrical conductivity the effective mass of conductivity in polycrystalline material can be expressed as (Bulat et al., 2010c)

$$m_c = \frac{4\,m_{c\,11}}{1 + \sqrt{1 + 8\,m_{c\,11}\,/\,m_{c\,33}}}\,,\qquad(14)$$

where $m_{c\,ii}$ are effective conductivity masses along main crystalline directions ($i = 1, 2, 3$).

The boundary scattering of holes was taken into account using relaxation time in the form $\tau_{b,h} = L_n\,/\,v$. The relaxation time energy dependence for acoustic scattering is $\tau_a \sim \varepsilon^{-1/2}$. It is the same as that for point defect scattering or alloy scattering in solid solution. It appears that this energy dependence is the same also for boundary scattering of holes. So the change of mobility in nanostructured material can be describe as (Bulat et al., 2010c)

$$u = \frac{L_n\,/\,l_s}{1 + L_n\,/\,l_s}\,u_s\,,\qquad(15)$$

where $l_s$ and $u_s$ are the mean free path and mobility in initial solid solution. The Lorenz number and the Seebeck coefficient in this case are the same as for acoustical scattering due to the same energy dependencies of the relaxation times.

The experimental values of electrical conductivity together with estimations based on equation (15) are shown on Fig. 20. In the initial solid solution $Bi_{0.4}Sb_{1.6}Te_3$ the values of electrical conductivity in the cleavage plane and the Seebeck coefficient were equal to $1000\ \Omega^{-1}cm^{-1}$ and $195\ \mu V/K$ correspondingly. In $Bi_{0.3}Sb_{1.7}Te_3$ these values were equal to $1387\ \Omega^{-1}cm^{-1}$ and $187\ \mu V/K$. The experimental values of mobility in $Bi_{0.3}Sb_{1.7}Te_3$ were 15% higher than in $Bi_{0.4}Sb_{1.6}Te_3$. The values of effective masses were taken from two different sources (Luk'yanova et al., 2010) and (Stordeur et al., 1988). The effective masses of the density of state per one ellipsoid $m_{d1}$ and of conductivity $m_c$ obtained using (14) were equal to $0.069m_0$ and $0.054m_0$ (Luk'yanova et al., 2010) and $0.305m_0$ and $0.186m_0$ (Stordeur et al., 1988). Due to the wide spread of the effective mass values the estimations of the mean free path in the initial solid solution $Bi_{0.4}Sb_{1.6}Te_3$ were quite different $l_a$=23 nm and 4nm correspondingly. This is reflected on the Fig. 20 where the effect of boundary scattering is more prominent for the estimations with larger $l_a$ (compare curves 1 and 1').

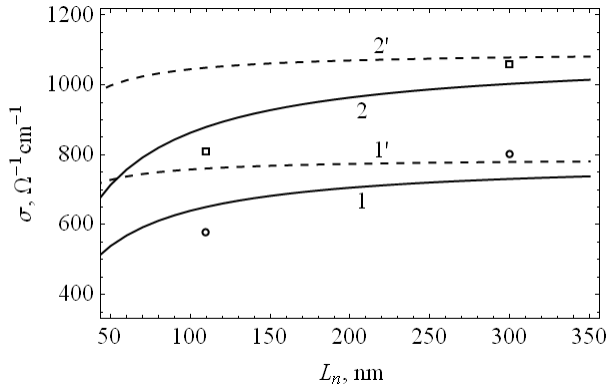

Fig. 20. The dependence of electrical conductivity of bulk nanostructured materials on the grain size for $Bi_{0.4}Sb_{1.6}Te_3$ (circles– experimental data; 1, 1′ – estimations) and for $Bi_{0.3}Sb_{1.7}Te_3$ (squares – experimental data; 2, 2′ – estimations). Estimations use effective mass values from (Luk'yanova et al., 2010) - 1, 2 and from (Stordeur et al., 1988) – 1′, 2′.

For estimations of the influence of boundary scattering on the phonon thermal conductivity the following material parameters were used. The lattice thermal conductivity in $Sb_2Te_3$ in the cleavage plane at room temperature is equal to $\kappa_{0,11}$ = 1.9 W/m K (Goltsman et al., 1972) and the anisotropy of the thermal conductivity is equal to 2.38 (Madelung et al., 1998). Averaging similar to (14) gives the thermal conductivity $\kappa_0$ = 1.47 W/m K. In $Bi_{0.4}Sb_{1.6}Te_3$ solid solution the thermal conductivity in the cleavage plane is 1.2 W/m K (Goltsman et al., 1972) that after averaging using anisotropy value of 2.22 (Madelung et al., 1998) gives $\kappa_s$ = 0.94 W/m K. The Debye temperature in $Sb_2Te_3$ is about 160K, Debye velocity was estimated as $v_D = 3.6 \cdot 10^5$ cm/s and the heat capacity at room temperature is close to usual value 24.9 J/mol K (Goltsman et al., 1972). This data allowed estimating the average mean free path in the pure crystal as $\bar{l}_0$ = 4.7 nm.

The comparison of the estimated thermal conductivity of nanostructured material with the experimental data is presented on Fig. 21. The electronic contribution to the thermal conductivity was subtracted using Lorenz factor calculated as described above. The results of estimations are quite well correlate with the experimental data. This allows one to conclude that the boundary scattering is important mechanism of reduction of phonon thermal conductivity in bulk nanostructured materials. The estimation of the decrease of the lattice thermal conductivity due to boundary scattering is shown on Fig. 22. It can be seen that the decrease can reach the values of 30-40% at the grain size of about 10-20 nm.

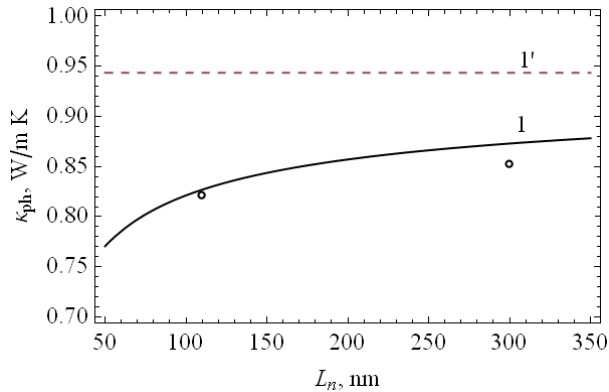

Fig. 21. The dependence of phonon thermal conductivity of bulk nanostructured materials on the grain size for $Bi_{0.4}Sb_{1.6}Te_3$ (circles – experimental data; 1 – estimations; 1' – the value in initial solid solution).

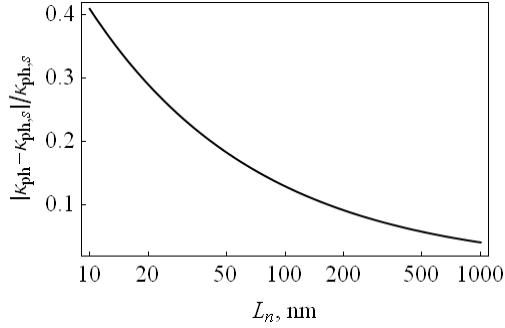

Fig. 22. The dependence of the relative decrease of lattice thermal conductivity on the grain size in nanostructured material based on $Bi_{0.4}Sb_{1.6}Te_3$ solid solution.

### 3.3 Energy filtering

In the previous section the influence of the boundary scattering on the electrical conductivity of nanostructured material was considered. The boundary scattering was described using constant mean free path equal to the size of grains $L_n$. As was noticed above the energy dependence of the relaxation time for boundary scattering in this approximation is the same as for acoustical scattering. So the Lorenz factor and the Seebeck coefficient should not differ from that in initial solid solution if the concentration remains the same. On the other hand the experimental data (Bulat et al., 2011a) showed the increase of the Seebeck coefficient in the samples with smaller grain size. To describe this effect the more detailed study of the scattering process was performed. The energy dependence of the probability of carrier scattering on the potential barrier at the grain boundary was taken into account. As the charge carriers with smaller energy scatter more intensively their contribution to the electrical and heat current decrease. This energy filtering can lead to the increase of the Seebeck coefficient if the energy relaxation length $l_\varepsilon$ is much greater than the momentum mean free path $l_p$ (Moizhes & Nemchinsky, 1998). At the temperatures much higher than Debye temperature the estimations for typical parameters of semiconductors (Moizhes & Nemchinsky, 1998) showed that as $l_p \sim 50$nm the energy relaxation length is about 500 nm that is much greater than the grain size considered in the present section.

There are several approaches that take into account the influence of the energy filtering on the transport coefficient. In (Ravich, 1995) the scattering on the single barrier was considered. In (Popescu et al., 2009) the exact expression for scattering probability was used but it was not taken into account that it should depend on the part of the kinetic energy corresponding to the motion normal to the boundary rather than the total energy. In (Mayadas & Shatzkes, 1970; Gridchin et al., 2005) the boundary scattering in polycrystalline thin films was considered but the relaxation time was anisotropic.

In the bulk nanostructured samples considered in this section the electrical conductivity appears to be isotropic due to random grain orientation and the following approach for calculation of relaxation time was used (Bulat et al., 2011a). In this approach the boundary scattering is modeled through the specular scattering on the randomly oriented planes representing grain boundaries and the inter plane distance is equal to the grain size $L_n$. The estimations of the mean free path in the previous section gave $l_p \sim 20$ nm. So the grain size is greater than the mean free path. In this case the multiple scattering can be taken into account through the summing up the probabilities of scattering rather than the matrix elements. In isotropic polycrystalline material with random grain orientation the summation of the probability of multiply scattering leads to the averaging over the boundary plane orientations. The total number of planes was estimated as $3L / L_n$, where $L$ is the characteristic sample size. Due to the conservation lows only two final states in the individual scattering act are possible, namely forward scattering and reflection. In the relaxation time calculation only the second type gives contribution. For the probability of reflection the exact expression is used $W_r(k_n) = 1 - D(k_n)$ where tunneling probability $D(k_n)$ is defined by equation (6) and $k_n$ is the wave vector normal to the grain boundary. As the number of incident electrons on the unit area of the boundary in one second is equal to the density of electron flow $j_i = \hbar k_n / mL$, the number of reflections in the unit time is equal to $j_i W_r(k_n)$. Finally the relaxation time can be calculated as

$$\tau_b^{-1} = \sum_{\mathbf{n}} \frac{\hbar k_n}{mL} W_r(k_n) \frac{-\Delta \mathbf{k_n} \, \mathbf{k}}{k^2},$$ (16)

where the summation over $\mathbf{n}$ takes into account all possible boundary orientations. The summation can be replaced with the integration over polar and azimuthal angles $\theta$ and $\phi$ determining the direction of normal vector $\mathbf{n}$. Then the expression for relaxation time due to boundary scattering can be obtained in the following form (Bulat et al., 2011a)

$$\tau_b^{-1} = \frac{6 \hbar k}{mL_n} \int_0^1 W_r(k\chi) \chi^3 \, d\chi,$$ (17)

where $\chi = \cos \theta$.

The experimental data and theoretical estimations for electrical conductivity and Seebeck coefficient in the bulk nanostructured materials based on $Bi_2Te_3$-$Sb_2Te_3$ solid solutions are presented on Fig. 23, 24. The scattering on the grain boundaries including energy filtering (17) and the scattering on acoustic phonons were taken into account. Because the exact account of anisotropy in $Bi_2Te_3$ based materials is complicated in equation (17) the density of state effective mass $m_{d1}$ was used. The unknown parameters in calculations were the width $d$ and the energy height $\varepsilon_b$ of the intergrain barrier. The estimations showed that quite good agreement with the experimental data can be obtained at the reasonable values of these parameters equal to $d = 5$ nm and $\varepsilon_b = 1.5 k_0 T$. The other parameters were the same as for the estimations of boundary scattering in the constant mean free path approximation discussed in the previous section. In order to check the applicability of relaxation time approximation the estimation of typical values of the relaxation time were made. It is known that relaxation time approximation is applicable if $\tau > \hbar / k_0 T$ and if the temperature difference on the length of mean free path is small compared to average temperature. If the temperature difference on the sample with $L \sim 0.1\,cm$ is 100K, then the temperature difference of the length of the order of mean free path about 10 nm is $10^{-3}$K. The estimations gave $\tau \sim 10^{-13}$ s and $\hbar / k_0 T = 2.5 \cdot 10^{-14}$ s at room temperature so the both criteria are well satisfied.

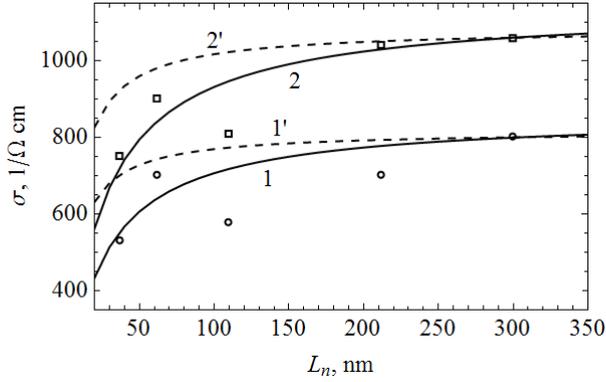

Fig. 23. The dependence of electrical conductivity of bulk nanostructured materials on the grain size for $Bi_{0.4}Sb_{1.6}Te_3$ (circles– experimental data; 1, 1' – estimations) and for $Bi_{0.3}Sb_{1.7}Te_3$ (squares – experimental data; 2, 2' – estimations). Estimations use effective mass values from (Luk'yanova et al., 2010) - 1, 2 and from (Stordeur et al., 1988) – 1', 2'.

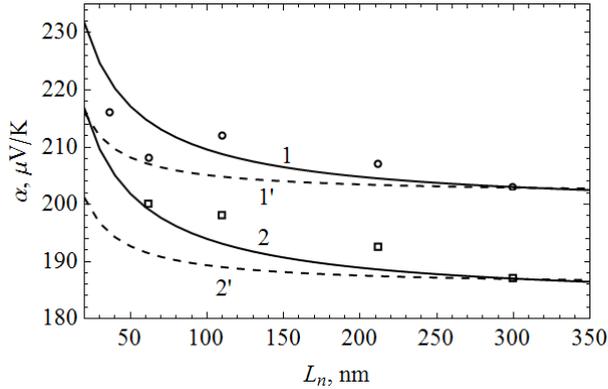

Fig. 24. The dependence of the Seebeck coefficient of bulk nanostructured materials on the grain size (see Fig. 23 for notation).

Finally the conclusion can be made that the energy filtering effect quite well describes the change of both electrical conductivity and the Seebeck coefficient in nanostructured materials. The estimations showed that in the bulk nanostructured materials based on $Bi_xSb_{1-x}Te_3$ the increase of the Seebeck coefficient due to this effect can reach 10-20% at the grain size of 20-30nm. If the lattice thermal conductivity decrease is the same as that for electrical conductivity this can give the 20-40% increase in the figure of merit.

## 4. Conclusion

The nanopowder p-Bi-Sb-Te with particles ~ 10 nm were fabricated by the mechanical activation method (ball milling) using different technological modes. Cold and hot pressing at different conditions and also SPS process were used for consolidation of the powder into a bulk nanostructure and nanocomposites.

Nanoparticles keep composition of initial solid solution $Bi_xSb_{2-x}Te_3$. The change of the hot pressing temperature did not result in the change of phase composition and of the lattice parameter of samples. The main factors allowing slowing-down of the growth of nanograins as a result of recrystallization are the reduction of the temperature and of the duration of the pressing, the increase of the pressure, as well as addition of small value additives (like $MoS_2$, thermally expanded graphite or fullerenes). The SPS processing is also an effective way for reduction of the CDA (or nanograins) size, and as consequence for improvement of the figure of merit. The best value of the efficiency ZT=1.12 (at the temperature ~ 350÷375 K) was measured in the hot pressed bulk nanostructures $Bi_{0.3}Sb_{1.7}Te_3$, while it was reached ZT=1.22 (at 360 K) in the bulk nanostructure $Bi_{0.4}Sb_{1.6}Te_3$ fabricated by SPS method.

The theoretical dependence of the electric and heat conductivities and the thermoelectric power as the function of nanograins size $L_n$ in $Bi_xSb_{2-x}Te_3$ bulk nanostructure are quite accurately correlates with the experimental data (see Sec. 3.2 & 3.3). It means that the phonons and holes scattering on the nanograin boundaries takes place. And the intensity of the scattering increases with reduction of $L_n$, that results in simultaneous decrease both phonons heat conductivity and electric conductivity. Our study shows that reduction of CDA size really can lead to improvement of the thermoelectric figure of merit.

Some theoretical results (Sec.3) on investigation of mechanisms of the thermoelectric efficiency improvement in bulk nanocrystalline semiconductors based on $Bi_xSb_{2-x}Te_3$ are summarized in Table 4.

The increase of the thermoelectric power by 10-20 % at $L_n$ =20 - 30 nm can lead to significant (20 - 40 %) increase of the thermoelectric efficiency provided that the reduction of the electric conductivity and the lattice heat conductivity compensate each other. In the investigated samples the full indemnification does not occur, however the thermoelectric efficiency nevertheless managed to be increased up to the values $ZT$ =1.1 - 1.2 (see Sec.2.6).

Table 4 shows that it is necessary to provide the small nanograin size (be more exact – CDA size) ~ 10 – 20 nm for realization of all three mechanisms of the figure of merit improvement. It is difficult to create such nanostructure technologically; the reason is the growth of the initial nanoparticles due to the recrystallization processes. However technological conditions have been determined (see Sec.2.5) for fabrication of the bulk nanostructures and nanocomposites based on $Bi_xSb_{2-x}Te_3$ solid solution from nanopowder by hot pressing and SPS methods which have given the reliable opportunity to obtain the CDA sizes $L_n$ ~ 40 nm.

| Mechanism of improvement Z | Ways of realization | Probable value of increasing Z |
|---|---|---|
| Additional phonon scattering | $L_n < (10 – 20)$ nm | (15 – 25)% |
| Tunneling of carriers | $L_n < (10 – 20)$ nm<br>Vacuum gaps between nanograins $\sim 1 – 2$ nm | ZT – up to 3,0 – 3,5 |
| Energy filtering of carriers | $L_n < (20 – 30)$ nm<br>Decrease of electrical conductivity and lattice thermal conductivity compensate each other | (20 – 40)% |

Table 4. Comparison of mechanisms of the figure of merit improvement

Fabrication of the vacuum gap $\sim 1 – 2$ nm between the nanograins for realization of the tunneling mechanism of the improvement of the figure of merit and the cutting off the phonons transport hardly will be possible by the technology of ball milling with the hot pressing or SPS process. Moreover, the electronic microscopy research has not found out any gaps between grains (or CDA) in studied nanostructures - no vacuum, no oxide (Sec. 2.4, 2.5). The accomplishment of all listed in Table 4 requirements to the structure of the nano-thermoelectrics based on $Bi_xSb_{2-x}Te_3$ solid solution should provide the increase of ZT up to 3,5 at the room temperatures. If the vacuum gaps $\sim (1 - 2)$ nm between the grains can not be created technologically, but if the bulk nanostructure with the grain sizes $\sim (10 - 20)$ nm can be realized, the increase of ZT up to 1.5 can be expected.

## 5. Acknowledgment

The work was supported by the Ministry of Education and Science of the Russian Federation, contract № 16.523.11.3002 and partially by the grant of the President of the Russian Federation № MK-7419.2010.2.